# EFFICIENT PHYSIOLOGICAL CONTROL OF AN INTEGRATED SYSTEM ARCHITECTURE FOR CONTINUOUS-FLOW VENTRICULAR ASSIST DEVICES: IN-SILICO STUDY

Bruno J. Santos[1]
brunojsantos@usp.br
Idágene A. Cestari[1,2]
idagene.cestari@incor.usp.br

[1]Programa de Pós-graduação em Engenharia Biomédica, Escola Politécnica, Universidade de São Paulo, São Paulo, SP, Brasil.
[2]Laboratório de Bioengenharia, LIM 65, Instituto do Coração, Hospital das Clinicas HCFMUSP, Faculdade de Medicina, Universidade de São Paulo, São Paulo, SP, Brasil.

Correspondent: Idágene A. Cestari,
idagene.cestari@incor.usp.br

**ABSTRACT**
This study presents the development and in silico evaluation of an Integrated System Architecture (ISA) for the physiological control of continuous-flow ventricular assist devices (VADs). The system employs an automatic controller based on pressure measurement at the VAD inflow cannula, enabling the estimation of heart rate and ventricular filling pressure and allowing dynamic speed adjustments to optimize VAD–patient interaction. The evaluation considered 27 simulation scenarios and four aspects of performance: (i) operating speed regulation, (ii) responsiveness to variable demand, (iii) mitigation of adverse events, and (iv) the impact on physiological variables. The tests performed confirmed that a) the controller adapted continuously to the preload, afterload, and heart rate; b) was capable of adaptation to flow demand with variations in the final operating speed between $-0.6 \cdot 10^3$ r/min (–10%) and $+2.6 \cdot 10^3$ r/min (+43%) relative to the initial speed of $6 \cdot 10^3$ r/min; c) resulted in no suction and backflow events, which were previously observed in 12 of the 27 scenarios, through dynamic speed adjustment. All adjustments were consistent with modifications of the physiological signals of pressure and volume or metabolic indicators. The results indicated an increased left ventricular ejection fraction (6.23–62.31% variation in most scenarios), reduction in ejection work (–1658.64 to –206.37 mmHg·ml), decreased pressure–volume loop area (–9485.17 to –1056.05 mmHg·ml), reduced myocardial oxygen consumption (–20.5 to –1.14 $mL \cdot min^{-1}$), increased oxygen delivery (+56.9 to +626.23 $mL \cdot min^{-1}$), and increased cardiac power (+100 to +3070 mW). Taken together, these findings demonstrate that the ISA for VAD physiological control can dynamically regulate speed, adapt to demand, mitigate adverse events, and influence hemodynamic and metabolic variables, establishing the basis for future "in vitro" validation.

## 1. INTRODUCTION

Cardiovascular diseases represent a critical global health challenge, with Congestive Heart Failure (CHF) being one of the most debilitating conditions. CHF is characterized by the heart's inability to pump sufficient blood to meet the body's metabolic demands, leading to a 5 years mortality rate of approximately 50% [1]. Cardiac transplantation is the gold standard treatment for advanced CHF; however, it is limited by donor scarcity and clinical contraindications [2].

In this context, Ventricular Assist Devices (VADs) have emerged as a cornerstone therapy, serving as a bridge-to-transplantation or destination therapy for ineligible patients [3]. These devices assist the mechanical function of the heart by supporting the left ventricle (LVAD) or right ventricle (RVAD). These continuous-flow devices, which operate on the principle of centrifugal pumps, significantly improve patient survival [4]. However, their long-term efficacy is compromised by high costs and complications, which frequently lead to hospital readmissions [5, 6].

A central limitation of commercial VADs is their operation at fixed speed settings. This configuration restricts the device's ability to dynamically adapt to variations in preload and afterload [7], thereby failing to emulate the Frank-Starling mechanism, which is essential for the physiological regulation of cardiac output [8]. The absence of automatic adjustment can lead to severe adverse events, including ventricular suction (excessive flow) and pulmonary congestion (insufficient flow) resulting from increased afterload [9, 10].

Therefore, the development of physiologic control strategies, which automatically adjust pump speed in response to the patient's hemodynamic needs, is imperative to optimize the interaction between the VAD and the native ventricle [11, 12].

This study investigated strategies to optimize the support provided by LVADs, aiming to achieve physiological synergy with the left ventricle (LV) while reducing the occurrence of adverse events associated with inadequate assistance from the device.

The main Research Question (RQ) is as follows: *How can a novel physiological control approach for LVADs be developed that operates in synergy with the LV, minimizes adverse events caused by inadequate support, aligns with state-of-the-art automatic control methods, and remains interpretable, straightforward to implement, and compatible with commercial VAD controllers?*

To address this overarching question, the following sub-questions were formulated:

- **RQ1:** Can the proposed approach operate under variable-speed conditions as opposed to conventional fixed-speed controllers?
- **RQ2:** Does the controller adequately respond to the patient's natural hemodynamic demands?
- **RQ3:** Can the system mitigate adverse events resulting from insufficient or excessive blood flow?
- **RQ4:** Does the proposed strategy lead to measurable improvements in clinically relevant physiological metrics?

Within this framework, we propose the development of a novel physiological control strategy based on dynamic speed variation, designed to appropriately adapt to changes in preload, heart rate (HR), and systemic vascular resistance (SVR), while mitigating the risk of ventricular suction and backflow (reflux). This approach provides automated control of VAD by being incorporated into an Integrated System Architecture (ISA-VAD), which leverages multimodal sensor-based monitoring and feedback estimation with operational redundancy.

The guiding hypothesis of this study is that, through a literature review and the use of a comprehensive computational simulator as the experimental environment, it is possible to design and validate an innovative physiological controller capable of consistently addressing the formulated RQs.

The main contribution of this study is the development of a novel physiological control strategy for VADs, which was implemented and evaluated using an in-silico hemodynamic simulation platform. This platform enabled an in-depth analysis of physiological scenarios and performance metrics that have not been previously explored in the literature, thereby expanding our understanding of the clinical impact of advanced control algorithms on VAD performance.

The remainder of this paper is organized as follows: Section II describes the related works of this approach; Section III describes the elements used in the development of the approach; Section III presents the results, which are discussed in Section IV, and the conclusion is presented in Section V.

## 2. RELATED WORKS

Schima et al. [13] conducted the first clinical evaluation of a physiological controller including fifteen patients in four settings: intensive care, general ward, rehabilitation cycling, and spirometry. This milestone study highlighted the potential benefits of physiological control for patients supported by VADs, namely, (i) a significant increase in pump flow during exercise, suggesting improved functional capacity; and (ii) subjective reports of enhanced comfort, indicating better quality of life.

In vitro experiments by Pauls et al. [14] compared physiological controllers and showed that approaches emulating the Frank–Starling mechanism via preload prevented suction and LV overload. This technical superiority was corroborated by Tchantchaleishvili et al. [9] and Petrou et al. [15].

Preload-based physiological control requires the measurement of pressure-related variables such as blood pressure. Recent advances have enabled the integration of pressure sensors into VADs [16]. Key pressure-feedback proposals for preload control include Bullister et al. [17], Stevens et al. [18], Stephens et al. [19], and Petrou et al. [20].

Bullister et al. [17] introduced a hierarchical controller. Level 1 (primary) regulates the LV end-diastolic pressure (EDP), maintaining it within a clinician-defined range and scaling the pump flow to the incoming blood volume, thereby preventing backflow and suction. Level 2 regulates mean arterial pressure (MAP) by adjusting the pump's mean outlet pressure to a dynamic target determined by HR, which is used as a proxy for increased pumping demand.

For biventricular support, Stevens et al. [18] proposed a master–slave scheme with two VADs, each equipped with two pressure sensors and one flow sensor. The master implements the Frank–Starling law by defining the flow as a function of LVEDP, whereas the slave tracks the master linearly.

Stephens et al. [19] presented a controller with a sigmoidal control curve analogous to Starling's, using LV/RV EDP feedback. The design increases the sensitivity at low-to-moderate preloads while attenuating it at high preloads to avoid an excessive flow. An additional mechanism emulates the venous return curve, promoting an automatic balance between pulmonary and systemic circulation.

Petrou et al. [20] adjusted pump speed using the peak pressure measured at the inflow cannula, interpreted as LV systolic pressure (SP) under two conditions: (i) preload changes produce corresponding changes in stroke volume (SV), cardiac output (CO), aortic pressure (AoP), and SP; and (ii) when preload and afterload increase simultaneously and filling remains below the plateau of the end-systolic pressure–volume relationship (ESPVR), SP reflects preload. The system prevents suction, overload, and backflow.

Building on this evidence, the ISA-VAD interface implements automatic flow control in response to demand via variations in: (i) HR, a direct indicator of pumping requirements, and (ii) preload, the Frank–Starling expression of venous return per cardiac cycle. The pump speed is adjusted proportionally to these variations, while accounting for the afterload (vascular resistance) to maintain the flow within the desired range.

Early detection and mitigation of adverse events in VAD recipients are pivotal for their quality of life and survival [21]. Consequently, recent proposals have incorporated safeguards against suction and backflow [14]. Vollkron et al. [22] advocated maintaining the flow pulsatility index (PI) to prevent suction, computed each cardiac cycle as the difference between maximum and minimum pump flow. Subsequent approaches [23, 24] also used PI: Choi et al. [23] estimated it from the hemodynamic ripple in the motor waveform, and Petrou et al. [24] derived it from the pressure waves at the inflow cannula.

To prevent backflow, Vollkron et al. [25] proposed the maintaining of a fixed minimum pump speed. Although simple, this can create a “dead zone” that mismatches the individual physiology. Petrou et al. [24] mitigated this by incorporating a minimum-flow threshold into a proportional–integral speed controller. Although useful in specific contexts [15], fixed-minimum strategies may underperform during prolonged support because they overlook interindividual physiological variability [26].

To address these limitations, Leao et al. [27] employed artificial intelligence (AI) and introduced a fuzzy logic controller that regulates the speed through heuristic rules, ensuring a safe minimum flow. Recent studies [28, 29] further support AI-based physiological controllers as a means of overcoming the constraints of linear control and the inherent nonlinearity of physiological systems. Nevertheless, clinical adoption remains under rigorous evaluation because of concerns regarding model safety, interpretability/explainability [30], and integration with commercial controllers [31].

Finally, Vollkron et al. [24] required a flow sensor, which limits its clinical applicability. Petrou et al. [20] addressed this with a pressure-based PI variant. However, neither strategy explicitly handles backflow events during support and both rely on a constant minimum value that narrows the operational range. To overcome this, the ISA-VAD incorporates suction and backflow detection/mitigation by continuously monitoring (i) a minimum threshold for the estimated flow, which, where feasible, triggers an immediate speed increase, and (ii) minimum thresholds for the pump motor current PI and the VAD inflow pressure, which, when reached, trigger, where feasible, a speed reduction.

To facilitate the analysis of the aforementioned approaches, Table I provides a comparative summary of the physiological controller strategies described in this study. The comparison ranges from the control philosophy and algorithmic strategy to the evaluation methods and techniques for mitigating suction.

TABLE I. COMPARATIVE SYNTHESIS OF PHYSIOLOGICAL CONTROL APPROACHES FOR VENTRICULAR ASSIST DEVICES (VADS) FROM RELATED WORKS.

| Physiological Controller | Control Philosophy | Algorithmic Strategy | Key Input Signals | Sensor Type | Suction Mitigation Strategy | Backflow Mitigation Strategy |
|---|---|---|---|---|---|---|
| **Bullister et al. [17]** | **Hierarchical regulation of preload and afterload.** | **Hierarchical integral control: Level 1 (LVEDP), Level 2 (MAP via HR).** | **Left Ventricular end-diastolic pressure (LVEDP), Heart rate (HR) & Mena Arterial Pressure (MAP).** | **Pressure** | **Maintenance of LVEDP within a pre-set, safe range.** | **Hierarchical control ensures adequate speed to prevent backflow.** |
| **Vollkron et al. [22, 25]** | **Demand-responsive flow** | **closed loop targets a set** | **HR & Pump flow** | **Flow** | **Adjustment based on flow** | **Maintaining a fixed minimum** |
| | **with suction prevention via an expert system.** | **flow (via HR and pulsatility); Expert system for suction (via flow);** | | | **pulsatility index (PI) calculated from maximum and minimum pump flow** | **pump speed** |
| **Stevens et al. [18]** | **Coordinated biventricular support through a master-slave approach to balance ventricular preloads.** | **Master defines flow via LVEDP (Starling-like); slave maintains a linear relationship between preloads.** | **LVEDP/RVEDP & pump flow.** | **Pressure & Flow** | **Coordination of pump speeds to maintain a balanced, linear relationship between LV and RV preloads, preventing volume shifts.** | **Master VAD's Starling-like control ensures adequate forward flow based on preload, preventing conditions for backflow.** |
| **Stephens et al. [19]** | **Independent Starling-type control for each ventricle to achieve automatic circulatory balance in biventricular support.** | **Sigmoid function using LVEDP/RVEDP feedback to control sensitivity.** | **LVEDP/RVEDP.** | **Pressure** | **High-sensitivity response to low preloads via sigmoid control curves and stable adjustment via a "system response path".** | **Starling-like increase in pump flow in response to higher preloads maintains robust forward propulsion.** |
| **Petrou et al. [20, 24]** | **Physiological adaptation by mimicking the Frank-Starling mechanism using LV systolic pressure as a preload indicator.** | **Proportional control based on peak inflow pressure.** | **LV Systolic Pressure.** | **Pressure** | **Adjustment based on PI derived from pressure waves at the inflow cannula** | **Minimum-flow threshold incorporated into a proportional–integral speed controller** |
| **ISA-VAD** | **Dual-Function Demand/Safety; automatic flow control in response to demand.** | **Proportional adjustment of pump speed to variations in HR and preload.** | **HR, Preload (via motor current PI and VAD inflow pressure).** | **Pressure** | **Continuous monitoring: minimum motor-current PI and inflow pressure trigger speed reduction** | **Continuous monitoring: minimum motor-current PI and inflow pressure trigger speed reduction** |

# 3. MATERIALS AND METHODS

## 3.1. Design of the physiological controller of the ISA-VAD

The ISA-VAD incorporates a physiological controller with a dynamic reference point to meet the patient's physiological demands based on the measurement of the pressure wave exerted on the VAD inlet cannula. After digital processing, this measurement indicated the patient's HR and preload values. The reference point was defined in terms of the outlet flow to compensate for variations in the afterload of the pump.

The physiological controller of the ISA-VAD employs a dynamic approach that adjusts the flow in real time according to variations in the patient's HR. The model was inspired by and adapted from the study by Vollkron et al. [25], which correlated the patient HR and assistance flow to optimize device performance. The continuous adaptation of the algorithm aims to maintain hemodynamic support amid rapid changes in oxygen demand, such as during exercise or at rest.

To model the relationship between the HR and VAD target flow, the central approach uses linear interpolation to estimate the intermediate values between two known points corresponding to pairs of HR and associated VAD flows.

In the algorithm, the rate of change between flow and HR, denoted as $\alpha HR$, was computed using (1), where $Flow_{upper}$ and $Flow_{lower}$ are the bounds of the VAD mean flow and $HR_{upper}$ and $HR_{lower}$ are the bounds of the patient's HR.

$$\alpha FC = \frac{Flow_{upper} - Flow_{lower}}{HR_{upper} - HR_{lower}} \quad (1)$$

The variable $\alpha HR$ represents the VAD flow change in response to changes in the patient's heart rate. With $\alpha HR$ computed, the algorithm determines the $desiredFlow_{HR}$ from the estimated $HR_{Estimated}$ in real time using (2).

$$desiredFlow_{HR} = Flow_{lower} + ((HR_{Estimated} - HR_{lower}) * \alpha HR) \quad (2)$$

The original approach by Vollkron et al. [25] assumes natural HR variability within a range of 60 to 120 beats per minute (beats/min), as the strategy relies on the relationship between HR and oxygen demand to adjust the VAD flow. In patients receiving beta-blockers, HR variability may be restricted, and the use of HR alone as the control variable may be limited because these drugs can reduce HR sensitivity to changes in physiological demand.

The algorithm implemented addresses the limitations observed in clinical practice, particularly in patients with reduced HR variability or hemodynamic conditions that require finer control. Additional hemodynamic demand indicators were incorporated to complement HR information. These indicators are based on estimates of LV preload derived from the VAD inlet pressure waveform.

Two complementary strategies were implemented to adjust the VAD target flow using the inlet pressure.

1. The VAD minimum inlet pressure ($IP_{min}$) value corresponds to the minimum LV filling pressure or minimum preload; this variable indicates end-diastolic LV blood volume (LV EDV) and guides VAD flow adjustment to avoid overload or inadequate filling. Specifically, monitoring $IP_{min}$ actively prevents ventricular suction events by triggering a proportional reduction in pump speed when abnormally low pressures indicate decreased preload and inadequate filling. Conversely, an elevated $IP_{min}$ indicates an accumulation of residual volume and potential ventricular overload, prompting the controller to increase VAD flow within safe operational limits to restore hemodynamic unloading.
2. The VAD maximum inlet pressure ($IP_{max}$) value corresponding to the maximum pressure observed during systole; this variable enables adjustment of the VAD flow to ensure support without competing with the flow generated by the patient's heart.

Combining these two adjustment points with HR analysis eliminates the dependence on HR.

Combining these two adjustment points with HR analysis reduces the strict reliance on HR as an isolated control variable. In scenarios of profound heart failure leading to a loss of native ventricular contraction or asystole (0 beats/min), the pulse pressure approaches zero ($IP_{max} \approx IP_{min}$). Under such boundary conditions, the controller is programmed to recognize the loss of synchronous pulsatility and gracefully default to a continuous, preload-sensitive baseline support mode, ensuring uninterrupted systemic perfusion. While the primary experimental scenarios evaluated physiological states between 60 and 120 beats/min, defining the 0 beats/min condition serves as a critical safety mechanism within the algorithm.

The integration of these hemodynamic variables into the algorithm supports flow control and prevents excessive or insufficient flow. The initial strategy associated $IP_{min}$ with preload estimation was based on the study by Stephens et al. [19], who used a sigmoidal function based on the inlet pressure. In this study, the minimum LV diastolic pressure was replaced by the minimum amplitude of the VAD inlet pressure waveform ($IP_{min}$) to simplify signal processing while preserving the relationship between the variable and demand.

To constrain the values within a defined interval and ensure more physiological transitions between the pressure limits, as expressed in (3), a sigmoidal function was employed. In this formulation, x denotes the input variable ($IP_{min}$), whereas A, B, and C are estimated parameters corresponding to the slope transition rate, inflection point (which determines the horizontal shift of the curve), and amplitude of the function.

$$f(x) = \frac{C}{1 + e^{-A*(x-B)}} \quad (3)$$

Table II shows the $IP_{min}$ (range 1–23 mmHg, per simulator operational range) values obtained under three HR conditions (60, 90, and 120 beats/min), together with the corresponding target flow values. In this control architecture, $IP_{min}$ and HR serve as independent input variables to the controller, whereas the mean VAD flow represents the resulting target setting (output) dictated by the algorithm.

These data were derived from simulations of a healthy cardiovascular system performed with the Computational Simulator of VAD Physiological Controllers (CS-PC-VAD) integrated with the Harvi Simulator, as previously described [32]. For each HR condition, sigmoidal parameters ($A$,$B$ and $C$) were estimated to represent the responses associated with $IP_{min}$. Parameter fitting was performed in MATLAB®(R2022b, MathWorks Inc.) using the Curve Fitter Toolbox via nonlinear least squares regression, minimizing the mean squared error between observed and model-predicted values. The application of the sigmoidal function in the control system enables (i) parameter adjustment to tailor the response curve to different conditions, (ii) gradual transitions to avoid abrupt responses to small $IP_{min}$ changes; and (iii) output value constraints within a defined range.

TABLE II. EXPERIMENTAL MINIMUM VALUES OF THE PRESSURE WAVE IN THE INFLOW CANNULA ($IP_{min}$) OF THE VENTRICULAR ASSIST DEVICES (VAD) AS A FUNCTION OF THE TARGET VAD FLOW AND HEART RATE (HR), ALONG WITH THE VARIABLES (A, B, AND C) OF THE SIGMOIDAL FUNCTION.

| Mean VAD Flow (L·min⁻¹) | $IP_{min}$ (mmHg) | Heart Rate (beats/min) | Heart Rate Classification | A | B | C |
|---|---|---|---|---|---|---|
| 3.27 | 1 | 60 | Low | 0.43 | 0.32 | 5.75 |
| 4.81 | 4 | | | | | |
| 5.38 | 7 | | | | | |
| 5.62 | 11 | | | | | |
| 5.73 | 14 | | | | | |
| 5.77 | 17 | | | | | |
| 5.8 | 19 | | | | | |
| 3.53 | 1 | 90 | Normal | 0.39 | 1.12 | 7.76 |
| 5.67 | 3 | | | | | |
| 6.77 | 7 | | | | | |
| 7.36 | 11 | | | | | |
| 7.69 | 15 | | | | | |
| 7.87 | 19 | | | | | |
| 8.01 | 22 | | | | | |
| 3.4 | 2 | 120 | High | 0.2 | 3.69 | 9.23 |
| 5.85 | 5 | | | | | |
| 7.28 | 11 | | | | | |
| 8.12 | 14 | | | | | |
| 8.64 | 18 | | | | | |
| 9 | 21 | | | | | |
| 9.26 | 23 | | | | | |

The final sigmoidal output was related linearly to $\alpha HR$ (1), parameterized to express the dependence of the variation in HR estimated across the HR lower and upper bounds.

Three HR ranges and corresponding expressions for *αHR* were defined as follows: low HR (< 80 beats/min) using (4); normal HR (80 ≤ HR ≤ 100 beats/min) using (5); and high HR (> 100 beats/min) using (6).

$$\alpha HR_{lowHR} = \frac{HR_{estimated}}{HR_{lower}} \quad (4)$$

$$\alpha HR_{normalHR} = \frac{2 * HR_{estimated}}{(HR_{upper} + HR_{lower})} \quad (5)$$

$$\alpha HR_{highHR} = \frac{HR_{estimated}}{HR_{upper}} \quad (6)$$

Figure 1 shows the sigmoidal equations fitted to the experimental data; the parameters were estimated according to the values reported in Table II. For low HR condition, the fitted sigmoidal parameters were *A*=0.43, *B*=0.32 and *C*=5.75; for normal HR, *A*=0.39, *B*=1.12 and *C*=7.76; and for high HR, *A*=0.20, *B*=3.69 and *C*=9.23. These equations characterize the behavior of the target flow across HR ranges, considering the variation of $IP_{min}$ over the interval 1–23 mmHg and the sensitivity coefficients *αHR* corresponding to the HR lower and upper bounds, set at 60 beats/min and 120 beats/min, respectively. The equations were organized into three groups corresponding to the following HR ranges:

1. The target flow for low HR (60 beats/min ≤ HR < 80 beats/min) as a function of $IP_{min}$ variation was as follows (7).

$$desiredFlow\ _{IPmin}^{lowHR} = \frac{5.8}{1 + e^{-0.43 * (IPmin - 0.32)}} * \alpha HR_{lowHR} \quad (7)$$

2. The target flow for normal HR (80 beats/min ≤ HR ≤ 100 beats/min) as a function of $IP_{min}$ variation was as follows (8).

$$desiredFlow_{IPmin}^{normalHR} = \frac{7.8}{1 + e^{-0.39 * (IPmin - 1.12)}} * \alpha HR_{normalHR} \quad (8)$$

3. The target flow for high HR (100 beats/min ≤ HR ≤ 120 beats/min) as a function of $IP_{min}$ variation was as follows (9).

$$desiredFlow_{IPmin}^{highHR} = \frac{9.2}{1 + e^{-0.2 * (IPmin - 3.7)}} * \alpha HR_{highHR} \quad (9)$$

It is important to acknowledge that some of the target VAD flows computed for high $IP_{min}$ configurations are notably elevated. The computational model in this study utilizes a generic continuous-flow centrifugal pump characteristic curve. However, clinically available devices (e.g., HeartMate 3) exhibit specific hydraulic limits and power constraints. Therefore, in practical translation, the physiological controller must integrate saturation limits to cap the target flow, ensuring that the commanded speed does not exceed the mechanical capabilities or safety thresholds of the specific implanted device.

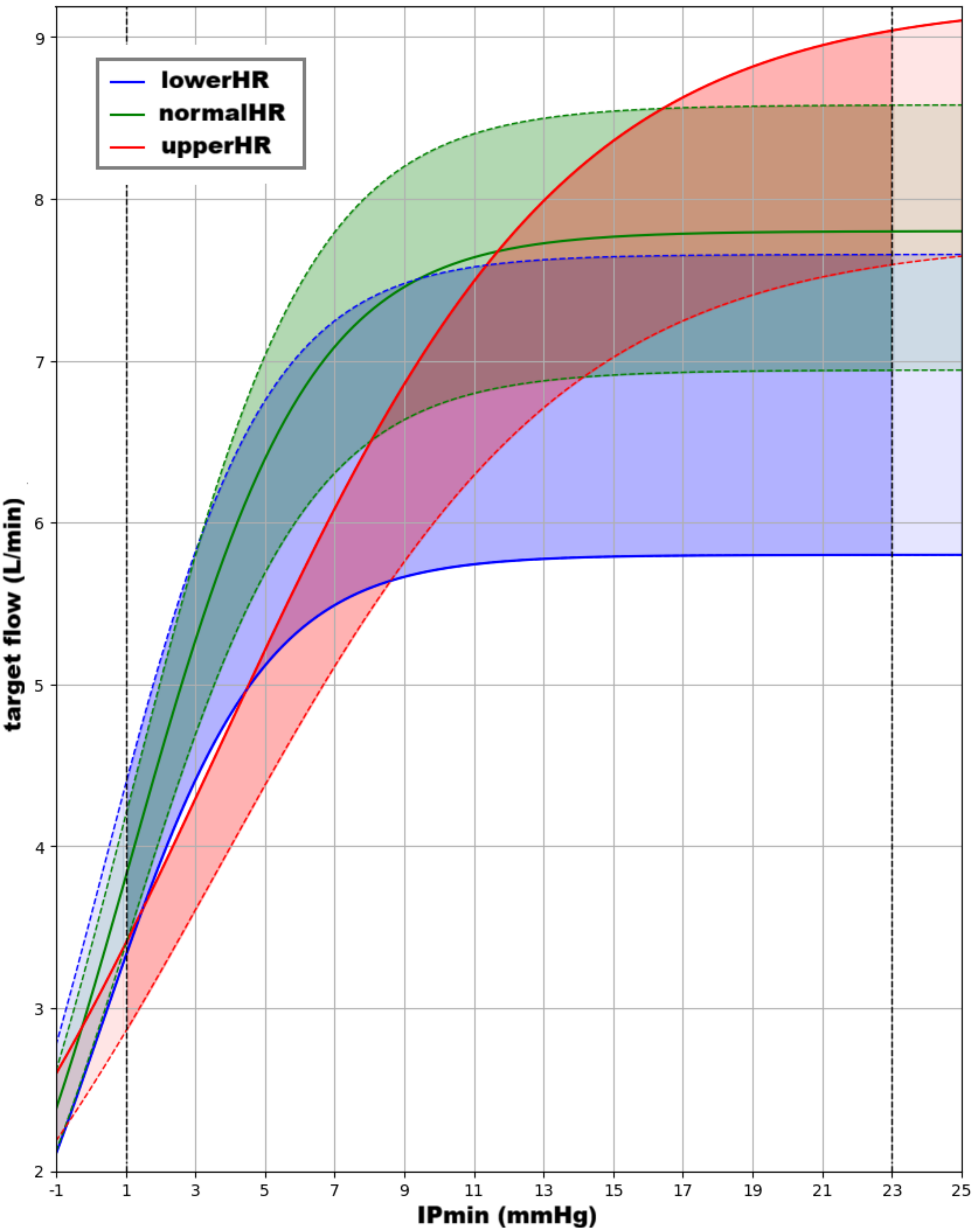

Fig. 1. Sigmoidal curves of the optimal flow as a function of heart rate (HR) ranges, considering IP_min values between 1 and 23 mmHg. The curves represent different HR classifications: low (blue), normal (green), and high (red). For low HR, the variation occurs between the basal value (αHR = 1) and above the basal value (αHR = 1.32). For normal HR, the range included values below (αHR = 0.89), at (αHR = 1), and above (αHR = 1.1) the basal level. For a high HR, the variation extended from below basal (αHR = 0.83) to the basal (αHR = 1).

The second strategy uses $IP_{max}$ as a preload indicator, in line with Petrou et al. [20], who estimated the preload from the VAD inlet pressure. Two modifications were implemented: (i) substitution of SP with the maximum VAD inlet pressure ($IP_{max}$); and (ii) replacement of proportional control with a sigmoidal function applied to $IP_{min}$. Applying a sigmoidal function to $IP_{max}$ variations enhanced the sensitivity to preload changes and reduced abrupt flow transitions.

Table III shows the $IP_{max}$ (range 63–185 mmHg, per simulator operational range) values obtained under three HR conditions (60, 90 and 120 beats/min), together the with corresponding target flow values. These data were derived from simulations of a healthy cardiovascular system performed with the CS-PC-VAD integrated with the Harvi Simulator, as previously described [32]. For each HR condition, sigmoidal parameters ($A$,$B$ and $C$) were estimated to represent responses associated with $IP_{max}$ divided by 10 to reduce the input magnitude. Fitting was performed in MATLAB® using the Curve Fitter Toolbox, as previously described.

TABLE III. EXPERIMENTAL MAXIMUM VALUES OF THE PRESSURE WAVE IN THE INFLOW CANNULA ($IP_{max}$) OF THE VENTRICULAR ASSIST DEVICE (VAD) AS A FUNCTION OF THE TARGET VAD FLOW AND HEART RATE (HR), ALONG WITH THE VARIABLES (A, B, AND C) OF THE SIGMOIDAL FUNCTION.

| **Mean VAD Flow (L·min⁻¹)** | **$IP_{max}$ *(mmHg)*** | ***Heart Rate (beats/min)*** | ***Heart Rate Classification*** | ***A*** | ***B*** | ***C*** |
|---|---|---|---|---|---|---|
| **3.27** | **76** | **60** | **Low** | **0.33** | **7.39** | **6.29** |
| **4.81** | **111** | | | | | |
| **5.38** | **125** | | | | | |
| **5.62** | **134** | | | | | |
| **5.73** | **140** | | | | | |
| **5.77** | **146** | | | | | |
| **5.8** | **152** | | | | | |
| **3.53** | **67** | **90** | **Normal** | **0.23** | **8.98** | **9.42** |
| **5.67** | **109** | | | | | |
| **6.77** | **131** | | | | | |
| **7.36** | **144** | | | | | |
| **7.69** | **154** | | | | | |
| **7.87** | **161** | | | | | |
| **8.01** | **168** | | | | | |
| **3.4** | **63** | **120** | **High** | **0.2** | **10.3** | **11.03** |
| **5.85** | **109** | | | | | |
| **7.28** | **136** | | | | | |
| **8.12** | **154** | | | | | |

| Mean VAD Flow (L·min⁻¹) | $IP_{max}$ (mmHg) | Heart Rate (beats/min) | Heart Rate Classification | A | B | C |
|---|---|---|---|---|---|---|
| 8.64 | 167 | | | | | |
| 9 | 177 | | | | | |
| 9.26 | 185 | | | | | |

Figure 2 shows the sigmoidal equations fitted to the experimental data; the parameters were estimated according to the values reported in Table III. For the low HR condition, the fitted sigmoidal parameters were *A*=0.33, *B*=7.39 and *C*=6.29; for normal HR, *A*=0.23, *B*=8.98 and *C*=9.42; and for high HR, *A*=0.20, *B*=10.3 and *C*=11.03. These equations characterize the behavior of the target flow across HR ranges, considering the variation of $IP_{max}$ over the interval 63–185 mmHg and the sensitivity coefficients *αHR* corresponding to the HR lower and upper bounds, set at 60 beats/min and 120 beats/min, respectively. As in the $IP_{min}$-based strategy, the equations were divided into three groups:

1. The target flow for low HR (60 beats/min ≤ HR < 80 beats/min) as a function of $IP_{max}$ variation was as follows (10).

$$desiredFlow\,_{IPmax}^{lowHR} = \frac{6.3}{1 + e^{-0.34 * \left(\left(\frac{PEmax}{10}\right) - 7.4\right)}} * \alpha HR_{lowHR} \quad (10)$$

2. The target flow for normal HR (80 beats/min ≤ HR ≤ 100 beats/min) as a function of $IP_{max}$ variation was as follows (11).

$$desiredFlow\,_{IPmax}^{normalHR} = \frac{9.42}{1 + e^{-0.23 * \left(\left(\frac{PEmax}{10}\right) - 9\right)}} * \alpha HR_{normalHR} \quad (11)$$

3. Thet target flow for high HR (100 beats/min ≤ HR ≤ 120 beats/min) as a function of $IP_{max}$ variation was as follows (12).

$$desiredFlow\,_{IPmax}^{highHR} = \frac{11.03}{1 + e^{-0.2 * \left(\left(\frac{PEmax}{10}\right) - 10.3\right)}} * \alpha HR_{highHR} \quad (12)$$

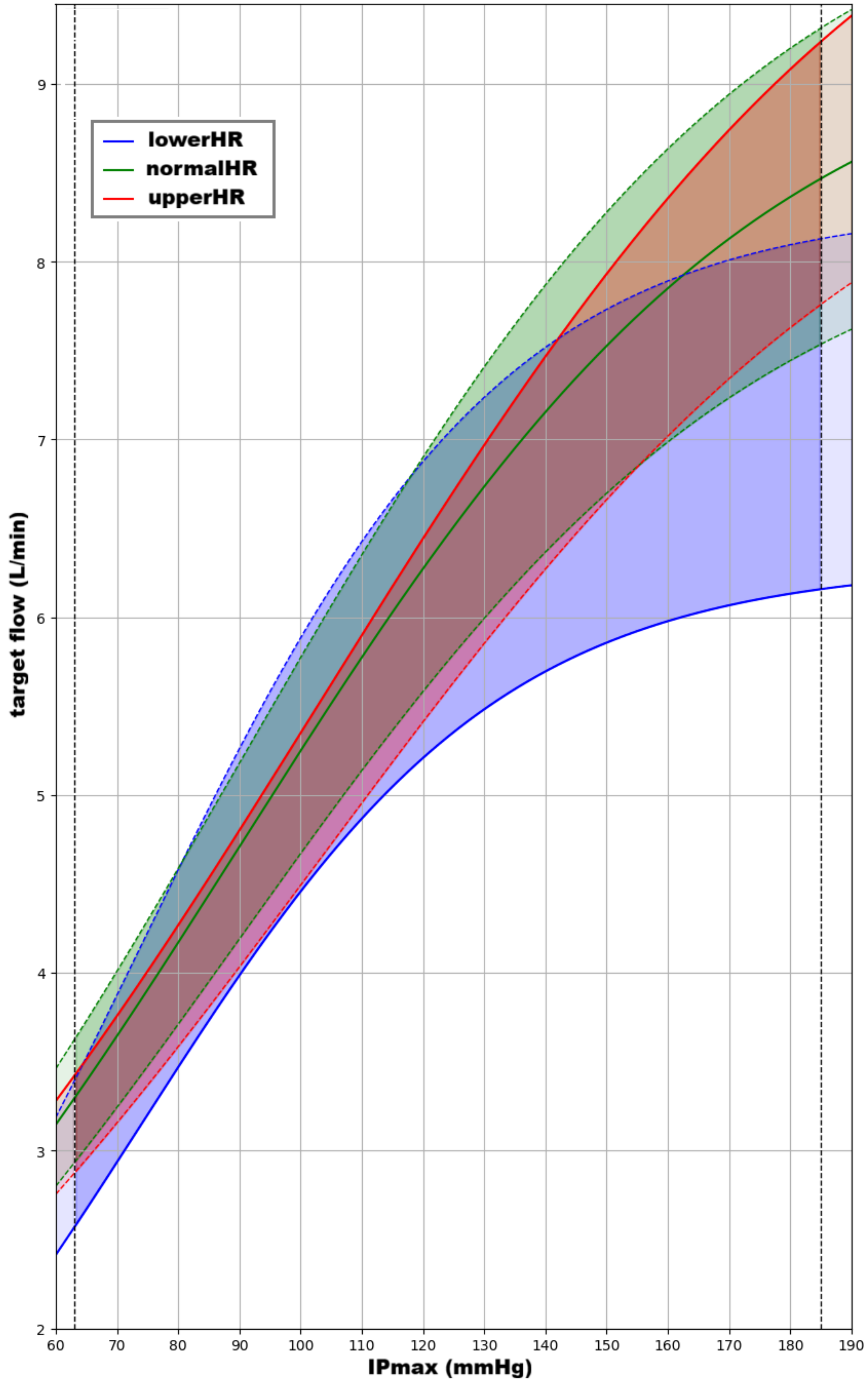

Fig. 2. Sigmoidal curves of the target flow as a function of the HR range, considering IPmax values between 63 and 185 mmHg. The curves correspond to the three HR classifications: low, normal and high. For low HR, the variation considers baseline (αHR=1.00) and above-baseline (αHR=1.32) values. For normal HR, the range included below-baseline (αHR=0.89), baseline (αHR=1.00) and above-baseline (αHR=1.10) values. For a high HR, the variation extended between below baseline (αHR=0.83) and baseline (αHR=1.00) values.

As illustrated in Figure 2 and Table III, the responsive target flow range driven by $IP_{max}$ in the high HR group (120 beats/min) is noticeably narrower compared to the low and normal HR groups. This behavior accurately reflects underlying physiological mechanics: at elevated HRs, the duration of diastole is disproportionately shortened, restricting the time available for ventricular filling. Consequently, the maximum achievable end-diastolic volume and the subsequent SP generation ($IP_{max}$) are blunted. The controller algorithm naturally accounts for this blunted contractility variance, resulting in a more constrained flow adjustment window under tachycardic conditions.

Fig. 1 and 2 display the curves fitted from the data in Table II and the parameters in Table III. The morphology differs from the classic sigmoidal "S" shape owing to the restricted operational range used in the analysis. A broader variable range would make the typical "S" structure more evident; nevertheless, the present focus is the sigmoidal relationship within the operational range of interest that reflects the practical operating limits of the system.

The ISA-VAD combines two flow adjustments ($desiredFlow_{Pmin}$ $e$ $desiredFlow_{Pmax}$) to obtain the desired final VAD flow ($desiredFlow_{FINAL}$). The combination is performed by a weighted arithmetic mean, with weighting factors $KP_{min}$ and $KP_{max}$, which adjust the relative influence of each point, as shown in (13).

$$desiredFlow_{FINAL} = \frac{(desiredFlow_{Pmin} * KP_{min}) + (desiredFlow_{Pmax} * KP_{max})}{2} \quad (13)$$

This approach allows the tuning of the relative contribution of $IP_{min}$ and $IP_{max}$ according to the patient or cardiovascular system conditions. Figure 3 shows the three-dimensional surfaces of target flow as functions of $KP_{min}$, $KP_{max}$, $IP_{min}$(1–23 mmHg), $IP_{max}$(63–185 mmHg) and coefficient $\alpha HR$.

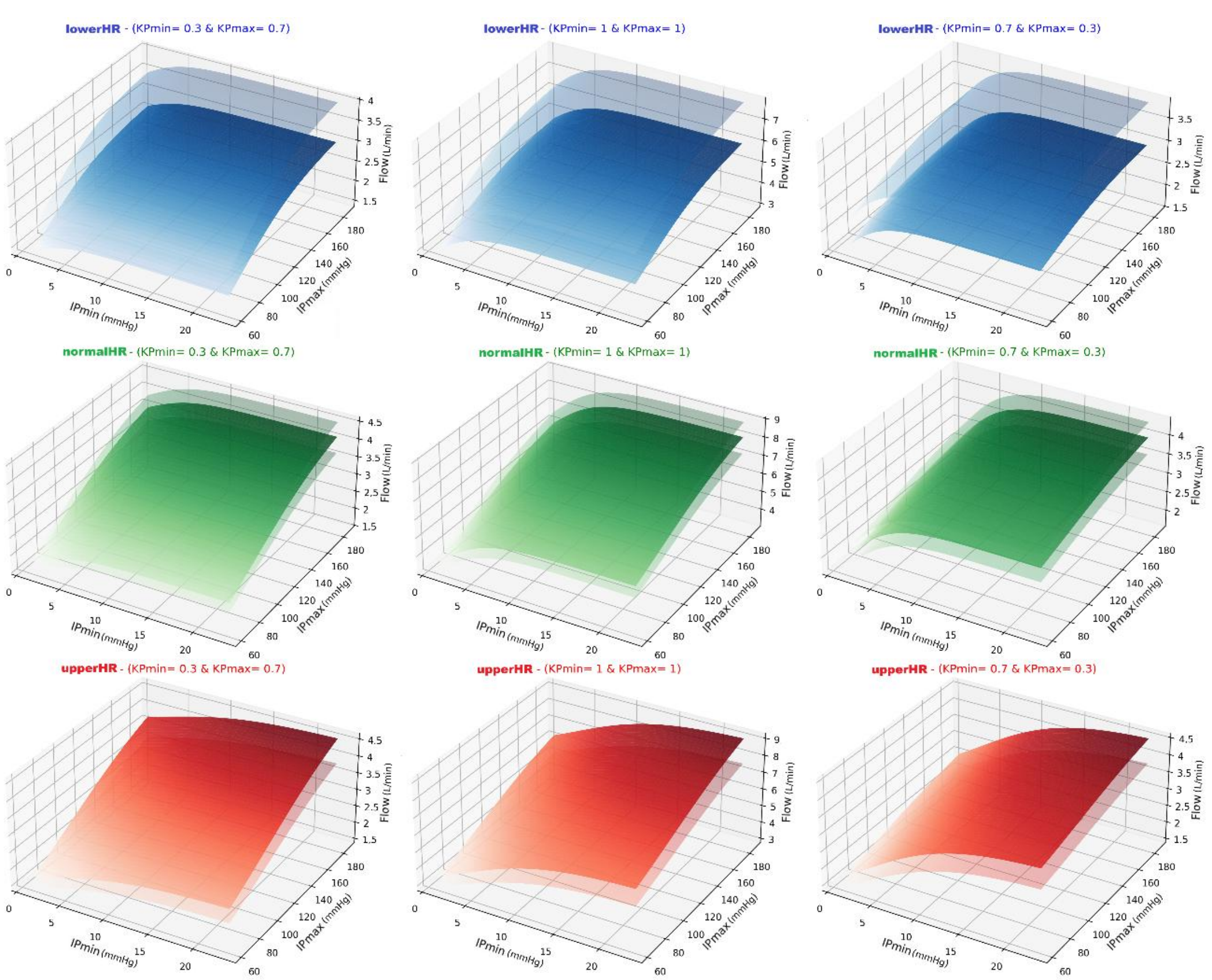

Fig. 3. Three-dimensional surface of the target flow curves by HR category, considering IPmin values between 1 and 23 mmHg and IPmax values between 63 and 185 mmHg, with variations in KPmin and KPmax.

Once the $desiredFlow_{FINAL}$ is determined, the VAD rotational speed ($desiredSpeed$) is computed by the direct proportion between $desiredFlow_{FINAL}$, the VAD maximum speed ($Speed_{upper}$), and the maximum allowed flow ($Flow_{upper}$), as expressed in (14).

$$desiredSpeed = desiredFlow_{FINAL} * \frac{Speed_{upper}}{Flow_{upper}} \quad (14)$$

The operational limits for the $desiredSpeed$ are the: minimum motor operating speed and maximum permitted speed. The VAD flow is constrained between the minimum and maximum VAD flows, consistent with the fluid resistance characteristics and device operating range.

$desiredSpeed$ was applied to a closed-loop control system. In the loop, the VAD output speed ($VAD_{outputSpeed}$) was continuously compared with $desiredSpeed$, producing an error input for the controller. A Proportional-Integral controller was implemented to correct the speed error. The proportional action adjusts the controller output based on the current error, whereas the integral action accumulates past errors to eliminate the steady-state deviations. The closed-loop Proportional-Integral control $VAD_{speed}$ formulation is defined in (15), where $KP_{speed}$ is the proportional gain, $KI_{speed}$ is the integral gain, $k$ denotes the current sample index, and $\Delta t$ denotes the sampling interval.

$$VAD_{speed}[k] = KP_{speed} * \left(desiredSpeed\,[k] - VAD_{outputSpeed}\,[k]\right) + KI_{speed} * \sum_{j=0}^{k} * \left(desiredSpeed\,[j] - VAD_{outputSpeed}\,[j]\right) * \Delta t \quad (15)$$

The ISA-VAD control system offers: (i) a weighted mean between $desiredFlow_{\,Pmin}$ $and$ $desiredFlow_{Pmax}$ to adjust the relative influence of each variable; and (ii) a closed-loop Proportional-Integral controller to reduce persistent errors and maintain performance under disturbances. The Proportional-Integral controller formulation was compatible with embedded system implementation.

### 3.2. Computational Implementation of the physiological controller of the ISA-VAD

This section presents the computational implementation of the ISA-VAD physiological controller using the CS-PC-VAD, as previously described [32].

The computational framework of this proposal comprises two primary components: (i) the generation of the dataset used for the in-silico simulation scenarios; and (ii) the implementation of the ISA-VAD physiological control algorithm. To elucidate the methodology, Figure 4 shows a flowchart that visually delineates the entire process. This diagram employs a standardized symbology, where processing stages are indicated in blue and decision conditions in yellow with arrows guiding the logical flow between steps. Each of these stages is examined in depth in the following sections.

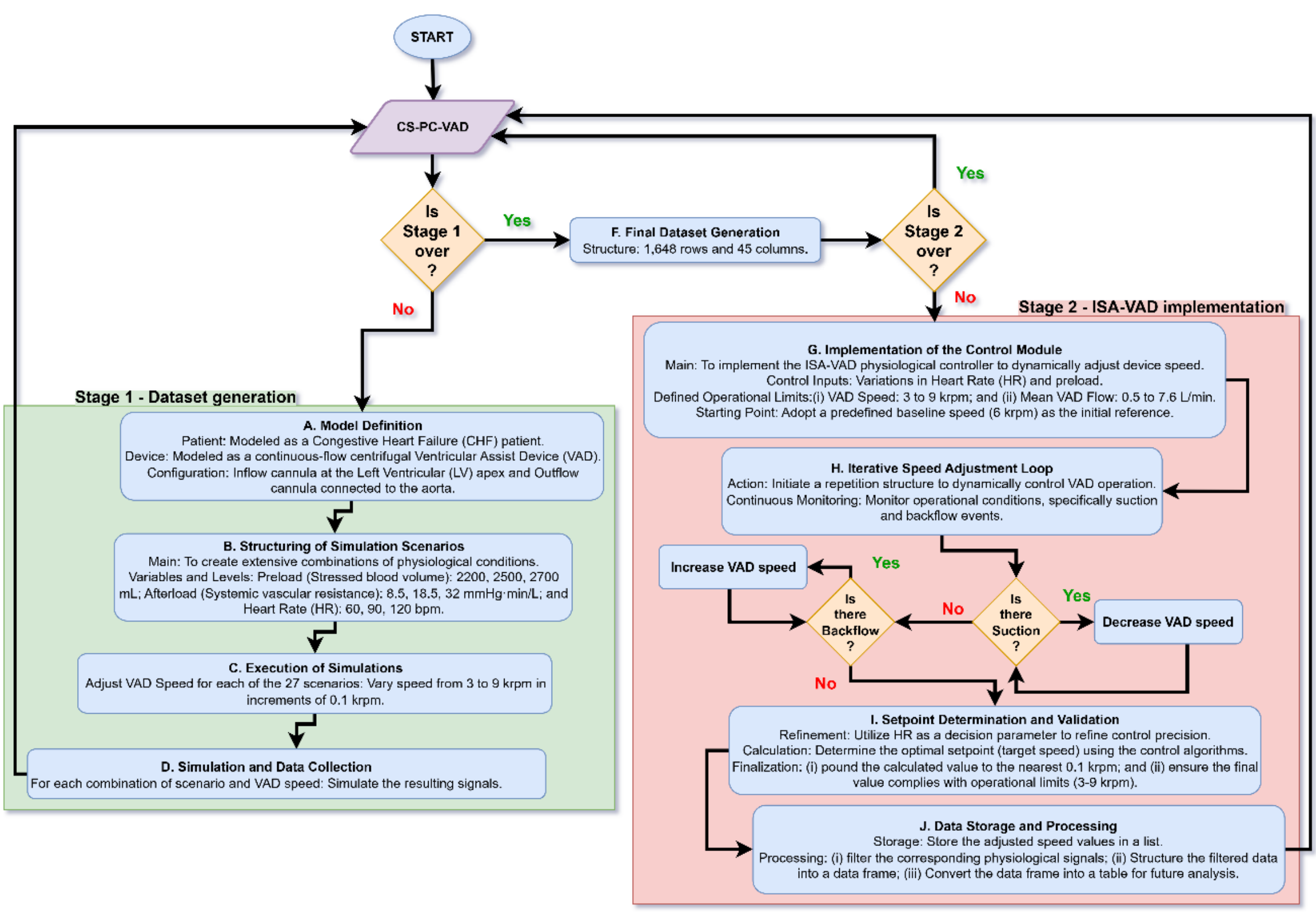

Fig. 4. Flowchart of the computational implementation and in-silico simulation of the ISA-VAD physiological controller. The flowchart depicts the process steps, using different colors to represent processing (blue) and decision-conditional (yellow) stages, with arrows indicating the sequence between them.

In Stage 1, the patient was modeled as a patient with CHF, and the device was modeled as a continuous-flow centrifugal VAD. The configuration included an inflow cannula placed at the LV apex and an outflow cannula connected to the aorta.

The simulations covered multiple physiological conditions corresponding to different CHF stages and activity levels, mitigating the limitations of linear models [29]. The scenarios were organized into 27 combinations of three variables: preload (stressed blood volume, mL), afterload (systemic vascular resistance, $mmHg \cdot min \cdot L^{-1}$), and HR (beats/min). The combinations were as follows.

- Scenarios 0–8: Afterload fixed at 8.5 $mmHg \cdot min \cdot L^{-1}$, with HR of 60, 90, and 120 beats/min and preloads of 2200, 2500, and 2700 mL.
- Scenarios 9–17: Afterload was increased to 18.5 $mmHg \cdot min \cdot L^{-1}$, with the same HR and preload variations.
- Scenarios 18–26: Afterload increased to 32 $mmHg \cdot min \cdot L^{-1}$, while, maintaining the same HR and preload variations.

This framework enabled the isolated assessment of the impact of each variable on the cardiovascular system while maintaining the other parameters constant. For instance, Scenarios 0, 3, and 6 were designed to investigate the effects of preload variation with heart rate fixed at 60 beats/min and constant afterload, whereas Scenarios 0, 9, and 18 allowed the analysis of afterload variation under fixed preload and heart rate conditions.

For each of the 27 simulation scenarios, the rotational speed of the VAD was adjusted from $3 \cdot 10^3$ r/min to $9 \cdot 10^3$ r/min in increments of $0.1 \cdot 10^3$ r/min. The resulting signals from these simulations were captured and stored as data-logs.

A comprehensive data wrangling process was applied to all 27 scenarios. The primary objective of this process was to transform the raw data, initially read from a spreadsheet file, into a structured, clean, and analysis-ready format for further analysis. Subsequently, a feature selection step was performed, creating a new DataFrame that retained only the columns of interest.

The dataset encompasses several parameter groups. The general physiological parameters included stressed blood volume as preload volume (mL), systemic vascular resistance as afterload ($mmHg \cdot min \cdot L^{-1}$), and HR (beats/min). The operational data for the VAD included the pump speed (rpm) and mean, minimum, and maximum blood flow rates ($L \cdot min^{-1}$). A significant portion of the variables were intravascular pressures (measured in mmHg), including systolic, diastolic, and mean values for the aorta, pulmonary artery, left atrium, and right atrium. Additionally, pulmonary capillary wedge pressure (PCWP) and central venous pressure (CVP) were included in the analysis.

Ventricular function metrics for both ventricles were detailed, comprising end-diastolic volume (EDV) (mL), end-systolic volume (ESV) (mL), EDP (mmHg), ESP (mmHg), SV (mL), CO ($L \cdot min^{-1}$), and LV Ejection Fraction (LVEF; %). Furthermore, cardiac efficiency indicators were incorporated, such as stroke work ($mmHg \cdot mL$), cardiac power (W), P-V area ($mmHg \cdot mL$), Myocardial Oxygen Consumption ($MVO_2$; $mL \cdot min^{-1}$) and Oxygen Delivery ($DO_2$; $mLO_2 \cdot min^{-1}$).

Specific data transformations were performed, including the conversion of a particular column to an integer data type. Subsequently, an initial exploratory data analysis was conducted to assess the data quality by identifying and quantifying null values within each column and across the entire dataset. This was complemented by the generation of a descriptive statistical summary, including the mean, standard deviation, and quartiles, to provide a quantitative overview of the processed data.

A descriptive statistical analysis of the dataset revealed an average LVAD flow of 4.8 $L \cdot min^{-1}$. However, the observation of minimum values reaching -2 $L \cdot min^{-1}$ and negative instantaneous averages suggests the potential occurrence of measurement artifacts or transient reverse flow. The incidence of ventricular suction was low (mean:

0.19), whereas the wide flow range (3.25-11.3 $L \cdot min^{-1}$) indicated significant variation in device performance.

Aortic pressures (systolic/diastolic/mean) were elevated, with averages of 115/107/111 mmHg, high variability (standard deviation > 40), and hypertensive peaks exceeding 265 mmHg. In contrast, pulmonary artery pressures (38/25/29 mmHg) exhibited a more stable distribution, although a subgroup of patients presented with pulmonary hypertension. Left atrial pressures show high variability, whereas right atrial pressures are more stable, ranging from 10–25 mmHg.

Volumetric analysis indicated severely compromised LV systolic function, with a mean EDV of approximately152 mL and a mean ESV of approximately126 mL, resulting in a low LVEF. Right ventricular (RV) function appeared to be preserved (mean EDV ~124 mL, mean ESV ~74 mL). LV EDP is highly variable (0-40 mmHg), reflecting a mixed population of patients with and without diastolic dysfunction.

The LV CO was very low (mean: 0.51 $L \cdot min^{-1}$, median: zero), indicating a strong dependency on mechanical circulatory support in these patients. Conversely, RV CO was preserved (~5.3 $L \cdot min^{-1}$), suggesting a compensated right-sided function. The mean LVEF was 33% and the mean RVEF was 45%, reinforcing the asymmetric ventricular dysfunction.

LV stroke work is high on average (~1,556 $mmHg \cdot mL$), but the presence of negative values indicates ventricular-suction failure. LV cardiac power was low (1.34 W), as expected in LVAD-dependent patients, whereas the RV power (0.33 W) was within the physiological range. The significant variability in $MVO_2$ and $DO_2$ reflects the heterogeneity of the study population.

Ultimately, the dataset generated from the 27 scenarios simulated within the CS-PC-VAD environment was structured into a final array of 1,648 rows and 45 columns.

In Stage 2, a structured program module with a physiological controller of the ISA-VAD was implemented in Python programming language within CS-PC-VAD to dynamically adjust the VAD rotational speed in response to variations in HR and preload, using the physiological controller of the ISA-VAD based on the algorithms described in Section "A. Design of the physiological controller of the ISA-VAD". The system's operational limits were set between 3 and $9 \cdot 10^3$ r/min, with the mean VAD flow constrained between 0.5 and 7.6 $L \cdot min^{-1}$.

During the physiological control adjustment, a predefined baseline speed was adopted as the initial reference point (speed of $6 \cdot 10^3$ r/min located at row 30 in the dataset). From this starting value, the device speed was iteratively adjusted using a repetition structure that dynamically controlled the VAD operation. The algorithm continuously monitored the operational conditions, specifically suction and backflow events. Upon detecting suction, the VAD speed decreased; conversely, if backflow was detected, the speed increased. HR was integrated as a decision parameter to refine the control precision and adapt the system response to the physiological variations. The optimal setpoint was determined using previously described control algorithms, and the resulting value was rounded to the nearest 100 rpm, while ensuring compliance with predefined operational limits.

The adjusted speed values were stored in a list, and the corresponding signals were filtered and structured into a data frame, which was subsequently converted into a table for further analyses. The results were visualized using time-series plots, enabling clear interpretation of the changes in VAD speed and associated physiological conditions.

A pseudocode algorithm is provided in Figure 5 to clarify the functionality of the structured program module responsible for the physiological controller of the ISA-VAD.

Procedure Main_Control
1. Read spreadsheet into a DataFrame named `data`.
2. Select columns.
3. Apply pre-processing.
4. Define initial thresholds:
  a. $Speed_{upper}$ = 9000;  b. $Speed_{lower}$= 3000; c. $Flow_{upper}$= 7.6; d. $Flow_{lower}$= 0.5;  e. $HR_{upper}$ = 120;  f. $Flow_{lower}$ = 60.
5. Initialize control variables:
  a. KP = 1,  b. is_finished = False, c. needs_read = True, d. is_physiological_phase = True.
6. Initialize data pointers and lists:
  a. row_number = 30; b. current_speed = data.loc[row_number, 'Speed (rpm)']; c. speed_list = [current_speed];
  d. initial_values_matrix = values from row row_number as an array
7. While is_finished is False do:
  a. If needs_read is True then:
    i. suction_state = (data.loc[row_number, 'Left_Ventricle_Suction (state)'] == 1);
    ii. reflux_state = (not suction_state) and (data.loc[row_number, 'LVAD_Min_Flow (L/min)'] < 0); iii. needs_read = False
  b. If suction_state then:
    i. row_number = row_number – 1;  ii. $Speed_{upper}$ = data.loc[row_number, 'Speed (rpm)']; iii. $Flow_{upper}$  = $(Speed_{upper} * Flow_{lower})$ / $Speed_{lower}$
    iv. Append upper_speed to speed_list
    v. needs_read = True
  c. Else if reflux_state then:
    i. row_number = row_number + 1;  ii. $Speed_{lower}$ = data.loc[row_number, 'Speed (rpm)']; iii. $Flow_{lower}$ = $(Speed_{lower} * Flow_{upper})$ / $Speed_{upper}$
    iv. Append lower_speed to speed_list
    v. needs_read = True
  d. Else if not is_physiological_phase then:
    i. is_finished = True
  e. If is_physiological_phase then:
    i. $HR_{Estimated}$ = data.loc[row_number, 'Heart_Rate (bpm)']
    ii. If $HR_{Estimated}$ < 80 then:
      $desiredFlow\ _{IPmin}^{lowHR}$ = Equation (**7**);  $desiredFlow\ _{IPmax}^{lowHR}$ = Equation (**10**)
    iii. Else if $HR_{Estimated}$ > 100 then:
      $desiredFlow\ _{IPmin}^{highlHR}$ = Equation ( **9** ); $desiredFlow\ _{IPmax}^{highlHR}$ = Equation (**12**)
    iv. Else:
      $desiredFlow\ _{IPmin}^{normalHR}$ = Equation (**8**);  $desiredFlow\ _{IPmax}^{normalHR}$ = Equation (**11**)
    v. $desiredFlow_{FINAL}$ = Equation (**13**)
    vi. $desiredSpeed$= $Speed_{lower}$ + $(desiredFlow_{FINAL}\ - Flow_{lower})$ * (($Speed_{upper}$- $Speed_{lower}$) / ($Flow_{upper}$ - $Flow_{lower}$))
    vii. final_number = round_up_to_nearest_hundred($desiredSpeed$)
    viii. final_number = clamp(final_number, $Speed_{lower}$, $Speed_{upper}$)
    ix. Append final_number to speed_list
    x. found_index = index in data where 'Speed (rpm)' == final_number
    xi. If found_index exists then row_number = clamp(found_index, 0, 60)
    xii. is_physiological_phase = False
    xiii. needs_read = True
8. scenario_values_matrix = values from row row_number as an array
9. result_dataframe = Create empty DataFrame
10. For each speed in speed_list do:
  a. filtered_rows = Filter data where 'Speed (rpm)' == speed; b. Concatenate filtered_rows into result_dataframe
11. result_matrix = result_dataframe converted to a matrix
End Procedure

Fig. 5. Pseudocode of the ISA-VAD physiological controller algorithm.

## 3.3. Computational Evaluation of the physiological controller of the ISA-VAD

The computational performance of the ISA-VAD physiological controller was evaluated through four in-silico experiments, each addressing one RQ. All tests were performed using the CS-PC-VAD integrated with a Harvi Simulator, as previously described [32].

Experiment 1 (E1) addressed RQ1 by comparing variable-speed control, implemented using the ISA-VAD physiological controller, to the standard constant-speed paradigm. Both control modes were tested in the CS-PC-VAD simulation environment, focusing on variations in the VAD operational speeds. The experiment comprised 27 scenarios, each starting with a predefined speed. The physiological controller then adjusted the VAD speed according to the inlet and outlet pressure conditions of the patient. The assistance values provided by the VAD were recorded for both modes and compared statistically.

Experiment 2 (E2) addressed RQ2 by evaluating the system's response to different flow demands. The aim was to test the ISA-VAD controller's ability to detect and regulate the VAD speed under changing demand conditions within CS-PC-VAD. As in E1, 27 scenarios were simulated. Each began with a standard VAD speed, after which the controller dynamically adjusted the parameters according to the pressure conditions. The primary outcome was VAD assistance, determined by the operational speed set by the ISA-VAD, and the values were compared across all scenarios.

Experiment 3 (E3) addressed RQ3 by testing the controller's response to adverse events, specifically backflow and ventricular suction. The analysis focused on speed variations generated by the ISA-VAD within the CS-PC-VAD system. A total of 27 scenarios were simulated, each beginning with predefined VAD speeds. The controller dynamically adjusted the speed by decreasing it when ventricular suction was detected and increasing it in response to the backflow. The resulting assistance levels, represented by the operational speed, were analyzed to quantify the response of the controller to these events.

Experiment 4 (E4) addressed RQ4 by assessing physiological outcomes. This study aimed to determine whether ISA-VAD dynamic speed adjustments improved simulated patient hemodynamics in the CS-PC-VAD system. Data from 27 scenarios were analyzed to evaluate key physiological metrics. These included LVEF, defined as the percentage of blood ejected from the LV per beat, with values below 25% indicative of CHF; LV work, representing the energy expended for blood pumping; and cardiac power, which reflects the total mechanical energy generated, partially compensated by the VAD. The LV P-V area was also examined as an indicator of the total energy per cardiac cycle, allowing a direct comparison of hemodynamic conditions before and after controller intervention. Finally, $MVO_2$ was analyzed as a measure of myocardial oxygen demand and $DO_2$ reflected the effective oxygen supply to the tissues.

**4. RESULTS**

This section presents the results of the in-silico evaluation, emphasizing the system's performance in four key aspects: (i) capacity to adjust operating speed, (ii) responsiveness to variable demand, (iii) ability to manage adverse events, and (iv) impact on simulated physiological variables.

Figure 6 shows at comparison between the initial and final states of the VAD operating speed, VAD flow, and patient pressure in different simulated scenarios. These results support the analyses of E1 and E2. The signals corresponding to the initial state are shown on the left, and those corresponding to the final state are shown on the right of the figure.

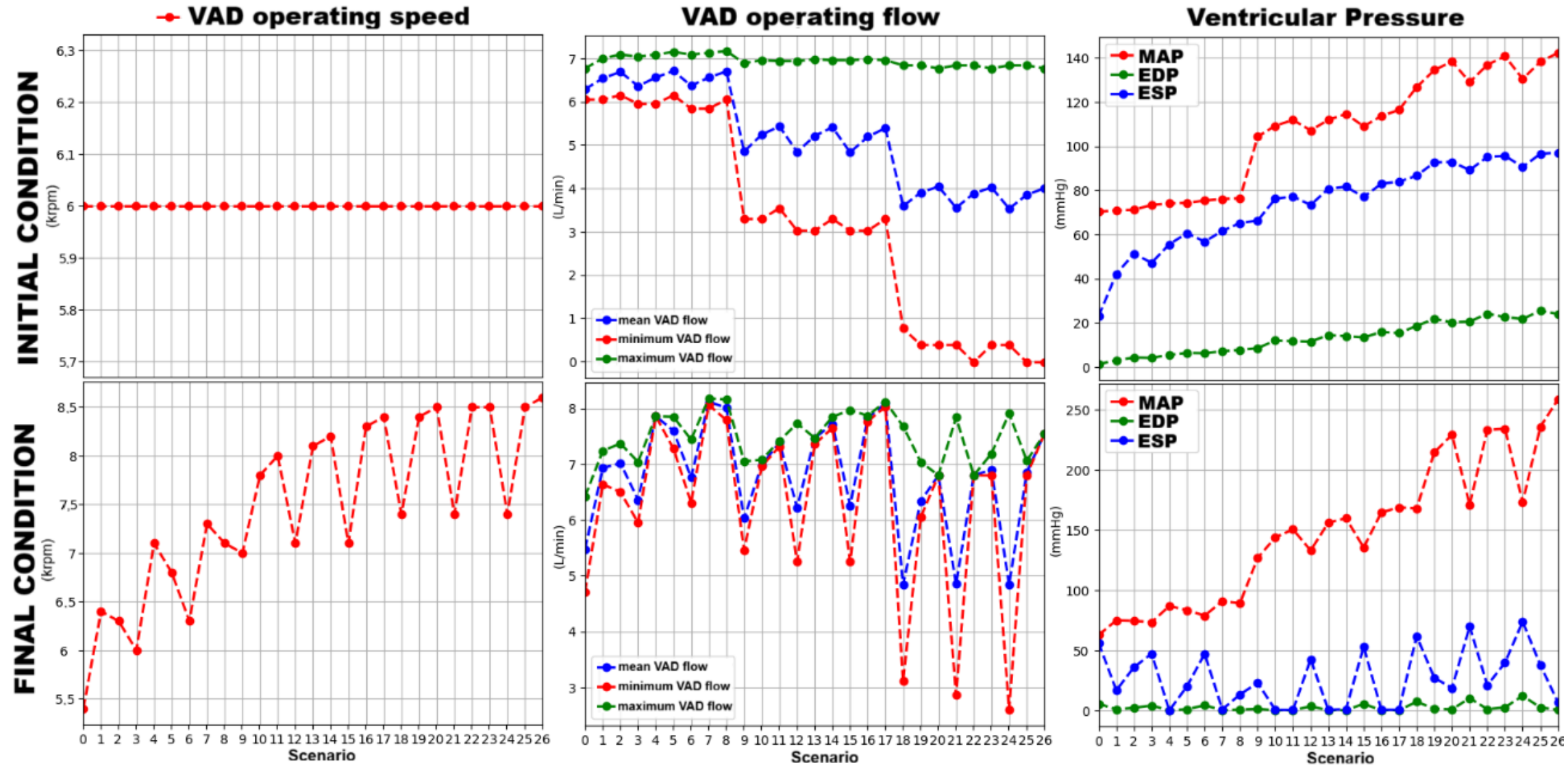

Fig. 6. Comparative analysis of a Ventricular Assist Device's (VAD) performance between initial and final states across 27 simulated scenarios, featuring instantaneous operating speed, output flow (mean, maximum and minimum), and key arterial pressures: Mean Arterial Pressure (MAP), End-Systolic Pressure (ESP), and End-Diastolic Pressure (EDP).

In Fig. 6, the left graphs display, in red, the VAD operating speeds recorded at the initial (upper) and final (lower) moments of each scenario; therefore, they do not represent the average variations over time within each scenario. The middle graphs correspond to the equivalent VAD output flow using the average (blue), maximum (green), and minimum (red) values. Finally, the right graphs present the equivalent pressures: ESP (blue), EDP (green) and MAP (red). All these pressures were obtained directly from the CS-PC-VAD.

It is crucial to address the hemodynamic discrepancy observed in the upper-right graph, where the MAP exceeds the native ESP, reaching unphysiologically high values (e.g., > 150 mmHg). In a native, unassisted cardiovascular system, MAP is mathematically bounded below the peak systolic pressure. However, under continuous-flow mechanical circulatory support, arterial hemodynamics become uncoupled from native ventricular systole. When the VAD operates at high rotational speeds against a fixed, elevated systemic vascular resistance, without the compensatory vasodilation typically provided by an intact baroreceptor reflex, the continuous pump output generates a sustained, non-pulsatile arterial pressure. Consequently, this device-driven MAP can easily surpass the diminished peak pressure generated by the failing native LV. This phenomenon explains both the unphysiological hypertension and the inverted MAP-to-ESP relationship observed in the severely impaired simulation boundary conditions.

The signals obtained indicate that, in response to E1, the physiological controller of the ISA-VAD does not operate at a fixed speed but rather adjusts its dynamics according to the different experimental scenarios. The empirical data corroborate this assertion, wherein the upper-left graph depicts a constant, standardized initial operating speed of $6 \cdot 10^3$ r/min across all 27 scenarios. In contrast, the lower-left graph reveals substantial variation in the final operating speed, demonstrating the controller's adaptive nature. These speed adjustments were a direct response to the hemodynamic conditions of each scenario, resulting in significant alterations in both VAD outflow (middle graphs) and systemic pressure (right graphs).

In response to E2, the system continuously adapts its operating speed according to the specific conditions of each process. This adaptation resulted in significant variations in the final VAD speeds, with the largest negative deviation at $-0.6 \cdot 10^3$ r/min (-10%) in Scenario 0, decreasing to $5.4 \cdot 10^3$ r/min relative to the standardized initial speed of $6 \cdot 10^3$ r/min, and the largest positive deviation recorded at $+2.6 \cdot 10^3$ r/min (+43.33%) in Scenario 26, reaching a peak of $8.6 \cdot 10^3$ r/min. These control actions profoundly impact hemodynamic parameters. An increase in the pump operating speed led to a corresponding increase in the MAP. This higher level of support also facilitates unloading of the ventricle, resulting in a reduction in preload.

Table IV shows a synthesis of the results related to the operating speed of the VAD with a physiological control strategy. Each of the 27 scenarios was characterized by three distinct speed parameters: initial, demand-driven average, and final safety-corrected speeds. The initial speed was consistently standardized to $6 \cdot 10^3$ r/min and served as the baseline for all simulations. The average operating speed, representing the controller's calculated response to meet physiological demand before safety interventions, varied across a wide range from a minimum of $5.4 \cdot 10^3$ r/min to a maximum of $9 \cdot 10^3$ r/min. This demonstrates the controller's capability to target a broad spectrum of support levels in response to different physiological contexts.

TABLE IV. SUMMARY OF OPERATING SPEED ADJUSTMENTS UNDER THE PHYSIOLOGICAL CONTROLLER OF THE ISA-VAD STRATEGY: COMPARISON BETWEEN A STANDARDIZED BASELINE DEMAND-DRIVEN TARGET AND FINAL SAFETY-MITIGATED SPEEDS ACROSS 27 SCENARIOS.

| Scenario | Initial ISA-VAD speed Adjustment ($\cdot 10^3$ r/min) | Intermediate ISA-VAD speed Adjustment ($\cdot 10^3$ r/min) | Initial detection of adverse events | Final ISA-VAD speed Adjustment ($\cdot 10^3$ r/min) | Final detection of adverse events |
|---|---|---|---|---|---|
| 0 | 6 | 5.4 | No | 5.4 | No |
| 1 | 6 | 6.4 | No | 6.4 | No |
| 2 | 6 | 6.3 | No | 6.3 | No |
| 3 | 6 | 6 | No | 6 | No |
| 4 | 6 | 7.1 | No | 7.1 | No |
| 5 | 6 | 6.8 | No | 6.8 | No |
| 6 | 6 | 6.3 | No | 6.3 | No |
| 7 | 6 | 7.3 | No | 7.3 | No |
| 8 | 6 | 7.1 | No | 7.1 | No |
| 9 | 6 | 7 | No | 7 | No |
| 10 | 6 | 8.3 | Suction | 7.8 | No |
| 11 | 6 | 8.5 | Suction | 8 | No |
| 12 | 6 | 7.1 | No | 7.1 | No |
| 13 | 6 | 8.4 | Suction | 8.1 | No |
| 14 | 6 | 8.7 | Suction | 8.2 | No |
| 15 | 6 | 7.1 | No | 7.1 | No |
| 16 | 6 | 8.5 | Suction | 8.3 | No |
| 17 | 6 | 8.9 | Suction | 8.4 | No |
| 18 | 6 | 7.4 | No | 7.4 | No |
| 19 | 6 | 8.9 | Suction | 8.4 | No |
| 20 | 6 | 9 | Suction | 8.5 | No |
| 21 | 6 | 7.4 | No | 7.4 | No |
| 22 | 6 | 8.9 | Suction | 8.5 | No |
| 23 | 6 | 9 | Suction | 8.5 | No |
| 24 | 6 | 7.4 | No | 7.4 | No |
| 25 | 6 | 9 | Suction | 8.5 | No |

| 26 | 6 | 9 | Suction | 8.6 | No |
|---|---|---|---|---|---|

The results in Table IV corresponding to E3 reveal the efficacy of the safety algorithm of the controller. Adverse events, specifically ventricular suction, were detected in 12 of the 27 simulated scenarios, invariably under conditions that required high pump speeds (>8.3 $\cdot$ $10^3$ r/min). The data clearly show the controller's corrective action: in every one of these 12 instances, the final operating speed was demonstrably lower than the initially targeted average speed. For example, in Scenario 20, a demand-driven speed of 9 $\cdot$ $10^3$ r/min triggered suction detection, prompting the controller to reduce the final operating speed to 8.5 $\cdot$ $10^3$ r/min. This speed adjustment represents a direct and successful intervention for mitigating adverse events. Notably, no backflow was detected in any of the scenarios.

Ultimately, the physiological controller of the ISA-VAD demonstrated a critical dual-function capability: it dynamically adjusted the speed to meet physiological requirements while simultaneously executing a safety override to prevent harm. By systematically reducing the pump speed upon detection of the suction risk, the controller effectively eliminated the conditions associated with this adverse event, thereby ensuring the safety and stability of the system operation across all simulated physiological challenges.

To address E4, physiological signals were collected for each of the 27 scenarios. Figure 7 shows the results obtained in simulation Scenario 9, characterized by the following variables: preload of 2200 mL, afterload of 18.5 $mmHg \cdot min \cdot L^{-1}$, and HR of 60 beats/min. In this context, preload represents the input volume, whereas the afterload corresponds to the resistance.

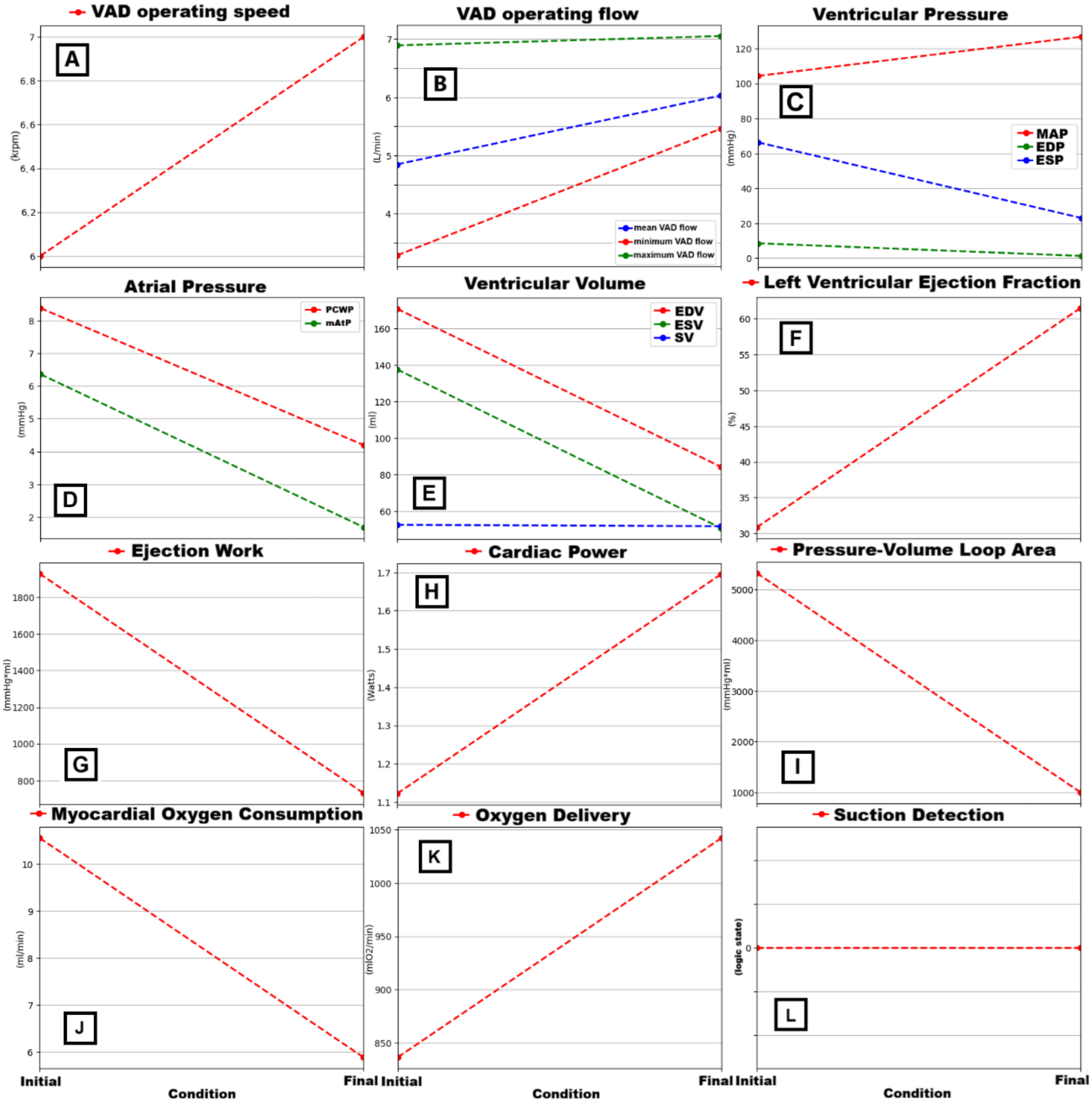

Fig. 7. Comparative analysis of the hemodynamic and metabolic outcomes of a Ventricular Assist Device (VAD) operation between the initial and final states in a simulated Scenario 9. The plots detail the following parameters: (a) VAD speed; (b) minimum, mean, and maximum VAD flow; (c) Ventricular Pressure with Mean Arterial Pressure (MAP), End-Systolic Pressure (ESP), and End-Diastolic Pressure (EDP); (d) Atrial Pressure with Pulmonary Capillary Wedge Pressure (PCWP) and Mean Atrial Pressure (mAtP); (e) Ventricular Volume with End-Diastolic Volume (EDV), End-Systolic Volume (ESV), and Stroke Volume (SV); (f) Left Ventricular Ejection Fraction (LVEF); (g) Ejection Work; (h) Cardiac Power; (i) pressure-volume (P-V) Loop Area; (j) Myocardial Oxygen Consumption ($MVO_2$); (k) Oxygen Delivery ($DO_2$); and (l) Suction Detection.

The results in Fig. 7 include: (i) adjustment of the operating speed (red line in Fig. 7a); (ii) VAD operating flow (minimum in red line, average in blue line, and maximum in green line in Fig. 7b); (iii) ventricular pressure (MAP in red line, EDP in green line, and ESP in blue line in Fig. 7c) and atrial pressure: PCWP in red line and mean atrial pressure (mAtP) in green line in Fig. 7d); (iv) ventricular volume (EDV in red line, ESV in green line, and SV in blue line in Fig. 7e); and (v) LVEF (red line in Fig. 7f). Metabolic and hemodynamic indicators: (i) Ejection Work (red line in Fig. 7g), (ii) Cardiac Power (red line in Fig. 7h), (iii) P-V loop area (red line in Fig. 7i), (iv) $MVO_2$ (red line in Fig. 7j), (v) $DO_2$ (red line in Fig. 7k), and (vi) Suction Detection (red line in Fig. 7l).

Fig. 7 demonstrates the physiological effects of the VAD operation in a simulated scenario under high preload and afterload conditions. The increase in the VAD operating speed and flow (Fig. 7a,b) resulted in left ventricular unloading. This unloading is quantified by the reduction in ventricular and atrial pressures (Fig. 7c,d) and the decrease in EDV and ESV (Fig. 7e). Consequently, native cardiac work was reduced, as measured by the decrease in ejection work, P-V loop area, and $MVO_2$ (Fig. 7g-j). Parallel to ventricular unloading, mechanical circulatory support increased cardiac power and $DO_2$ (Fig. 7h,k), indicating the optimization of systemic perfusion. Additionally, LVEF increased (Fig. 7f), which is consistent with the greater contractile efficiency under assistance. The operation of the VAD remained stable, with no suction events being detected (Fig. 7l).

To evaluate the efficacy of the ISA-VAD physiological controller, several cardiac performance and metabolic efficiency metrics were analyzed. Analysis of 27 operational scenarios of the ISA-VAD physiological controller demonstrated an increase in the mean rotational speed from $6 \cdot 10^3$ r/min to $7 \pm 0.89 \cdot 10^3$ r/min. This increment directly resulted in a corresponding increase in the mean blood flow from $5 \pm 1.16$ $L \cdot min^{-1}$ to $7 \pm 0.98$ $L \cdot min^{-1}$. Notably, the flow range was optimized, with the mean minimum flow increasing from $3 \pm 2.38$ $L \cdot min^{-1}$ to $7 \pm 1.55$ $L \cdot min^{-1}$, whereas the mean maximum flow remained stable at 7 $L \cdot min^{-1}$.

Hemodynamically, the MAP showed a marked elevation from 112 ± 26.15 mmHg to 151 ±59.47 mmHg, indicating improved systemic perfusion. Concurrently, the cardiac filling pressure was drastically reduced, indicating effective cardiac decompression. The mAtP decreased from 12 ±6.88 mmHg to 2 ±2.16 mmHg, and the PCWP declined from a mean of 14 ±6.48 mmHg to 5 ±1.84 mmHg.

The LV mechanics directly respond to this load adjustment. Preload was substantially reduced, as evidenced by the decrease in mean EDV from 192 ±35.36 mL to 77 ±52.94 mL and the fall in mean EDP from 13 ±7.41 mmHg to 1 ±3.19 mmHg. Ventricular emptying was also optimized, with the mean ESV reduced from 159 ±38.42 mL to 47 ±47.42 mL, accompanied by a decrease in the mean ESP from 77 ±19.21 mmHg to 21 ±23.73 mmHg.

Despite the significant reductions in ventricular volumes, the ventricular SV remained relatively stable, with the mean value changing from 47 ±7.2 mL to 45 ±15.06 mL. However, contractile efficiency, assessed by LVEF, showed a remarkable improvement, with the mean increase from 24 ±10.3 % to 61 ±17.47 %. Cardiac workload was significantly alleviated. This is reflected in the fall of the mean Ejection Work from 1709 ±376.67 mmHg·mL to 591 ±748.27 mmHg·mL and the drastic reduction of the P-V loop area, representing the net external systolic work, from a mean of 6883 ±2507.59 mmHg·mL to 819 ±1925.56 mmHg·mL. In contrast, the mean Cardiac Power increased from 1150 ±119.57 mW to 1860 ±947.22 mW.

Finally, the metabolic and oxygenation profiles confirmed the efficacy of this support. The reduced cardiac work led to a 50% decrease in mean $MVO_2$, from 18 ±6.87 $mL \cdot min^{-1}$ to 9 ±2 $mL \cdot min^{-1}$. Simultaneously, the enhanced circulatory output promoted an increase in systemic $DO_2$, with the mean value rising from 899 ±200.66 $mLO_2 \cdot min^{-1}$ to 1195 ±171.24 $mLO_2 \cdot min^{-1}$. This combined profile of reduced myocardial metabolic demand and increased global oxygen supply characterizes a successful intervention by the physiological controller of the ISA-VAD.

Figure 8 shows a comparative analysis of the results obtained in the 27 scenarios before (initial condition) and during assistance (final condition). Specifically, it demonstrates the effects on LVEF (Fig.8a), ejection work (Fig.8b), P-V loop area (Fig.8c), $MVO_2$ (Fig.8d), $DO_2$ (Fig.8e), and cardiac power (Fig.8f).

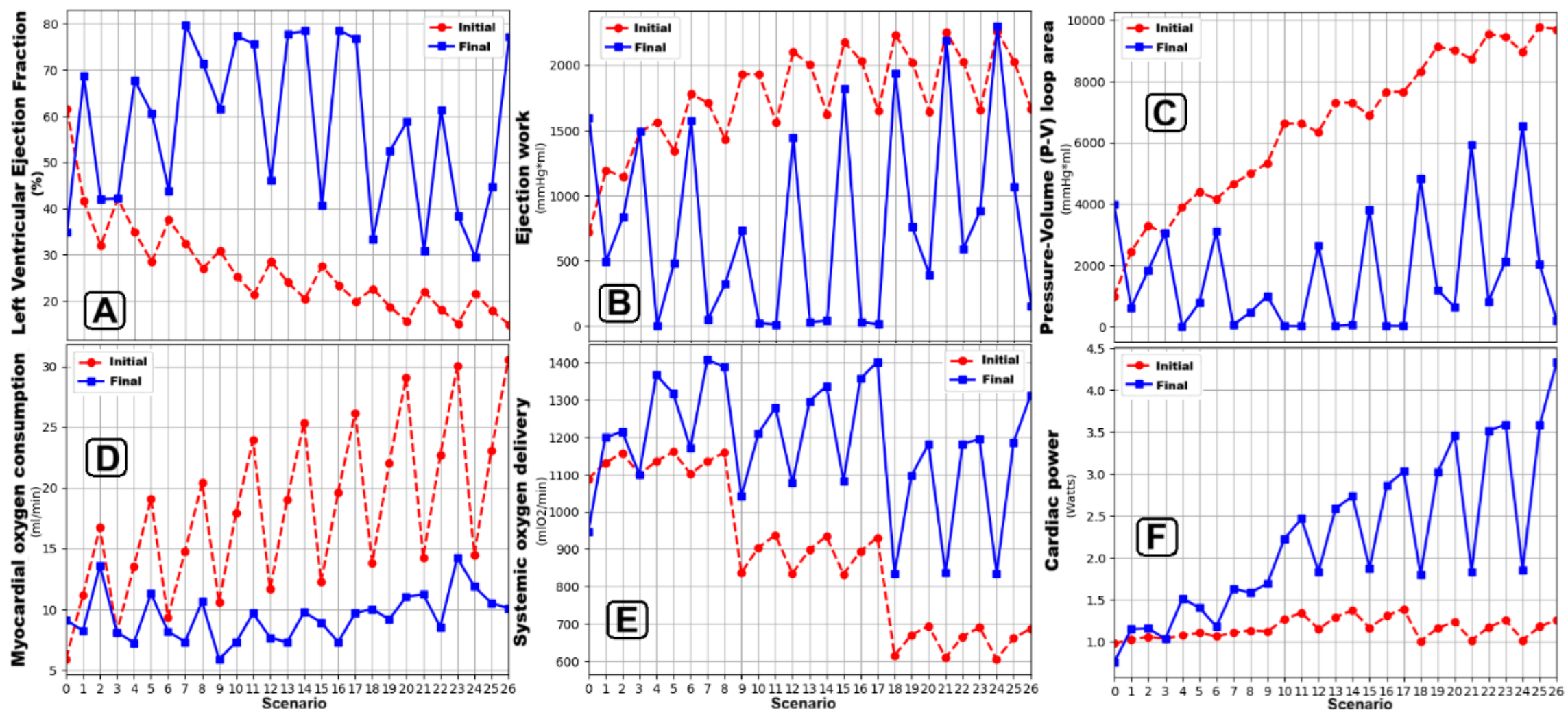

Fig. 8. Comparative analysis of the results obtained in the 27 scenarios before (initial condition) and during assistance (final condition): (a) Left Ventricular Ejection Fraction (LVEF), (b) ejection work, (c) Pressure-Volume (P-V) loop area, (d) myocardial oxygen consumption (MVO2), (e) systemic oxygen delivery (DO2), and (f) cardiac power.

Fig. 8a shows a comparison of the LVEF values between the initial and final conditions for the 27 simulated cases. Following activation of the ISA-VAD controller, an increase in the LVEF was observed in the vast majority of scenarios. The two notable exceptions were: (i) Scenario 0, representing a healthy subject (HR = 60 beats/min, preload = 2200 mL, afterload = 8.5 mmHg·min·$L^{-1}$), where LVEF decreased, consistent with the lack of need for mechanical support, and (ii) the baseline Scenario 3. Post-intervention, approximately 26% of the scenarios achieved the physiological LVEF range (50-70%), while 22% exhibited supranormal values. This finding underscores the importance of implementing safeguards in the control system to prevent excessive ventricular unloading. Excluding Scenarios 0 and 3, the controller's action regulated the LVEF from 6.23% to 62.31%, demonstrating its ability to adapt to diverse hemodynamic conditions.

The analysis of LV ejection work, as shown in Fig. 8b, revealed a significant reduction in myocardial work in over 85% of the simulations. This effect is attributable to the ventricular unloading provided by the ISA-VAD physiological controller, which assumes part of the pumping function. Disregarding Scenarios 0 and 3, the magnitude of this reduction ranged from -1658.64 mmHg·mL to -206.37 mmHg·mL. This range reflects the system's capacity to adaptively control LV load in response to the specific profile of each scenario, thereby optimizing the management of patients with CHF.

Fig. 8c shows the analysis of the LV P-V loop area, which represents the external systolic work. A reduction in this area was observed in most scenarios, indicating a lower mechanical energy expended per cardiac cycle. With the exclusion of Scenarios 0 and 3, the maximum and minimum reductions in the P-V area were -9485.17 and -1056.05 mmHg·mL, respectively. This effect is a direct consequence of decreased systolic pressure and volume resulting from the hemodynamic support provided by the ISA-VAD system.

Fig. 8d shows the impact of assistance on $MVO_2$. A generalized reduction in LV $MVO_2$ was verified, stemming from the alleviation of the cardiac workload. Excluding Scenarios 0 and 3, the largest drop in $MVO_2$ was -20.5 $mL \cdot min^{-1}$, while the smallest was -1.14 $mL \cdot min^{-1}$. These data corroborate the potential of the system to decrease myocardial metabolic stress, as the VAD performs a portion of the mechanical work, thereby reducing the oxygen demand for ventricular contraction.

The assessment of systemic $DO_2$ (Fig. 8e) showed, that in most cases, an increase occurred following the ISA-VAD intervention. Disregarding Scenarios 0 and 3, the incremental values of oxygen delivery ranged from 56.9 $mLO_2 \cdot min^{-1}$ to 626.23 $mLO_2 \cdot min^{-1}$. This benefit is directly associated with the optimization of cardiac output and consequent improvement in tissue perfusion, reinforcing the efficacy of ISA-VAD as a circulatory support strategy.

Finally, Fig. 8f shows the effect on LV cardiac power. The action of the ISA-VAD resulted in a general increase in power in most scenarios. Excluding Scenarios 0 and 3, the power increments were within the range of 100-3070 mW. This power enhancement is multifactorial, deriving from synergistic mechanisms, such as (i) Workload Reduction, which optimizes the heart's energetic efficiency (Fig. 8d); (ii) Preload Reduction, which improves contractile efficiency by operating at a more favorable point on the Frank-Starling curve (Fig. 6); and (iii) Augmented Total Aortic Outflow, generated by the VAD contribution, which increases the systemic ejected volume per cycle (Fig. 6).

## 5. DISCUSSION

The literature on continuous-flow VADs describes multiple physiological control strategies based on preload, afterload, and heart rate, aimed at optimizing the device-patient interaction [7, 9, 11, 16]. In this context, the ISA-VAD implements a system that dynamically adjusts the pump speed in response to hemodynamic changes, actively preventing adverse events such as ventricular suction and backflow.

The in-silico results from ISA-VAD in E1 demonstrated unequivocal adaptive behavior. Starting from a standardized initial speed of $6 \cdot 10^3$ r/min, the system adjusted its operation, resulting in final speeds that varied between 5.4 and $9 \cdot 10^3$ r/min across 27 hemodynamic scenarios. This result confirms that the ISA-VAD physiological controller does not operate in a fixed-speed mode but alters its rotation according to the imposed hemodynamic conditions.

The approach by Vollkron et al. [25] is intrinsically limited by the completeness and precision of the predefined expert rules. In contrast, the ISA-VAD physiological controller in E2 demonstrated a dynamic adjustment capability that transcended a fixed set of rules. The variation in the final operating speed (from $5.4 \cdot 10^3$ r/min to $8.6 \cdot 10^3$ r/min) across the 27 simulated scenarios (Fig. 6) corroborates that the controller operated according to an adaptive philosophy. This approach is conceptually similar to that proposed by Leão et al. [27], based on fuzzy logic and explicit expert knowledge, representing a top-down control model, as proposed by Fetanat et al. [28] and Son et al. [33] with Model-Free Adaptive Control (MFAC). The ISA-VAD controller does not maintain a constant speed; instead, it continuously alters its output in response to inferred hemodynamic conditions (variable demand), which is a central feature of MFAC systems that learn the dynamics of the system on-the-fly without an explicit mathematical model.

A challenge in the physiological control of VADs is finding a balance between maximizing cardiac output and preventing adverse events, such as ventricular suction.

The most conservative strategy is to maintain the cardiovascular system within a predefined "safe corridor". The controller proposed by Fetanat et al. [28], for example, operates to maintain the estimated preload within a safe range (e.g., 3–15 mmHg). Similarly, the controller proposed by Son et al. [33] used the LV EDP for the same purpose. Starling-like controllers by Bakouri et al. [34] and Wang et al. [35] geometrically implemented this concept by defining a "safe area" on a plot of pump flow versus preload surrogate. The objective was to maintain the operating point within this zone. In this philosophy, safety is an emergent property of the control law, and the primary task is to confine the system to safe limits, thereby ensuring an inherently stable operating state.

A different approach was adopted by Sadatieh et al. [36] with their Extreme Seeking Control (ESC). Instead of avoiding danger zones, this controller actively seeks the optimal performance point, which is defined as the pump speed immediately before the onset of suction. Analogously, controllers based on Deep Reinforcement Learning (DRL) adopted by Li et al. [37] learn the "safe zone" as an emergent property of their training, where the reward function penalizes states of suction or congestion. These approaches aim to extract the maximum performance by operating continuously at the failure threshold, which offers a minimal safety margin. Here, safety is not a zone to be maintained, but a boundary to be explored.

The ISA-VAD physiological controller presents a third approach that incorporates a dedicated "safety module" that combines structured algorithms and AI models to mitigate suction and backflow events, demonstrating accuracies of 99.7% and 98.4%, respectively, in experimental in-silico results [38]. The controller's results in E3 demonstrate a more conservative philosophy: in 12 of the 27 scenarios, in which the physiological demand required high speeds ($\geq 8.3 \cdot 10^3$ r/min), the safety algorithm detected a risk of suction and actively intervened, reducing the final speed (Table IV). For example, in Scenario 20, the target speed of $9 \cdot 10^3$ r/min was overridden and reduced to $8.5 \cdot 10^3$ r/min.

This hierarchical control architecture, in which a primary safety objective (avoiding suction) overrides a secondary performance objective (meeting flow demand), aligns the ISA-VAD precisely with the approach described by Fetanat et al. [28]. Both systems treat the maintenance of safety variables as the primary constraint. This modular approach allows safety to be an independent supervisory function rather than an integrated property of the control laws. Such separation is a critical advantage in medical device engineering, aligning with the principles of safety-critical systems and allowing robustness to be evaluated across heterogeneous libraries of cardiovascular models, in which failure modes vary with the underlying physiology [39].

Habigt et al. [40] demonstrated that the Power Ratio (PR) controller increases device support in response to an increase in the contractility of the native heart. This is because it was designed to add a fixed ratio of hydraulic power relative to the LV power output. Conversely, the Preload Responsive Speed (PRS) controller reduces its support as native contractility increases, because its objective is to add work to reach a fixed target curve. Therefore, as the heart contributes more work, VAD assistance decreases to maintain a constant target.

The intervention of the ISA-VAD in E4 resulted in a drastic reduction in cardiac workload, quantitatively evidenced by the drop in mean Ejection Work (from 1709 mmHg·mL to 591 mmHg·mL), in the P-V loop area (from 6883 mmHg·mL to 819 mmHg·mL), and, crucially, in $MVO_2$, which decreased by 50% (from 18 $mL \cdot min^{-1}$ to 9 $mL \cdot min^{-1}$). Simultaneously, the system improved overall hemodynamic performance: the mean pump flow increased from 5 $L \cdot min^{-1}$ to 7 $L \cdot min^{-1}$, MAP increased from 112 mmHg to 151 mmHg and $DO_2$ increased from 899 $mLO_2 \cdot min^{-1}$ to 1195 $mLO_2 \cdot min^{-1}$.

The significant increase in the mean LVEF from 24% to 61% suggests that the optimization of loading conditions (preload reduction, with a drop in EDV from 192 mL to 77 mL) allowed the native ventricle to operate at a more efficient point on the Frank-Starling curve. The mean increase in output (5-7 $L \cdot min^{-1}$) and $DO_2$, concomitant with the increase in mean speed ($6\text{-}7 \pm 0.89 \cdot 10^3$ r/min), indicates that the ISA-VAD enhances systemic perfusion on demand through a dual-mode philosophy. Initially, it functions synergistically, increasing its hydraulic power contribution relative to the heart's output, a principle similar to that of PR controllers [41]. Conversely, as native contractility recovers, the device exhibits a restorative function by reducing the assistance. This responsive weaning capability aligns with the operational design of PRS controllers [42].

The results confirmed that the ISA-VAD continuously adapts to the preload, afterload, and heart rate, thereby ensuring adequate perfusion and reducing adverse events. Improvements in LVEF, reduction in the P-V loop area, and optimized oxygen balance reinforce the potential of this approach to enhance ventricular support while decreasing the cardiac workload.

However, 22% of the scenarios presented supranormal LVEF, indicating a risk of excessive ventricular unloading. The literature highlights the complementary methodological advances.

The DRL controller proposed by Li et al. [37], which employs the soft actor-critic (SAC) algorithm, demonstrated superior performance in comparative experiments. The system exhibited a response time of only 38.6% of that of a Proportional-Integral-Derivative controller and reduced the Sum of Absolute Error (SAE) to 47.6% of the value obtained with the Proportional-Integral-Derivative, resulting in a faster and more effective response to the diverse physiological needs of patients.

In a complementary approach, Zapico et al. [43] introduced a DRL controller based on the Proximal Policy Optimization (PPO) algorithm. This controller was specifically designed to manage competing objectives, namely the prevention of ventricular suction and promotion of aortic valve opening. Compared with a constant-speed VAD operation, this controller ensured a more stable EDV (standard deviation of 5 mL vs. 9 mL) and a higher mean aortic flow (1.1 $L \cdot min^{-1}$ vs. 0.9 $L \cdot min^{-1}$).

Magkoutas et al. [44] presented an advancement in the data-driven physiological iterative learning controller (PDD-ILC). This controller is designed to achieve a physiological, pulsatile, and treatment-oriented response by utilizing counter-pulsation modes to minimize LV systolic work (LVSW) and co-pulsation modes to maximize pulsatility. The results demonstrated that the counterpulsation mode was capable of reducing LVSW by more than 50% compared to constant-speed support, while the copulsation mode dramatically increased aortic pulse pressure, directly addressing the limitations of continuous-flow VADs.

In contrast to the real-time adaptation approaches of DRL algorithms, the Genetic Algorithm Optimization Framework (GAOF) proposed by Magkoutas et al. [29] offers a secondary level of adaptation by personalizing the control parameters prior to clinical application. The GAOF optimizes complex control structures based on specific pump, patient, and therapy characteristics, considering the individual's pathology and the VAD model being used. This enables the development of customized, treatment-specific VAD controllers to improve clinical outcomes.

Within the framework of adaptive control, the parameters required to generate the sigmoid curves of the ISA-VAD physiological controller can be derived directly from the patient's clinical evaluations. Alternatively, they can be acquired through simulations specifically designed to mimic a particular patient, in alignment with the Health 4.0 concept [45].

Finally, the ISA-VAD architecture faces critical barriers in the adoption of chronic sensors owing to the risk of thrombosis, calibration drift, and system complexity. Although proxies derived from pump operational data are promising and less invasive alternatives, experimental results have demonstrated that physical sensors have superior performance in vitro [46]. This suggests that the future of this field may lie in hybrid approaches that combine intelligent estimators with optimized sensors to ensure both clinical reliability and long-term viability.

However, these future perspectives must be contextualized within the current translational limitations of this study. A major constraint is that the proposed simulation framework is based on a control measurement that is not yet available in standard clinical settings. Consequently, translating this setup into an implantable VAD involves significant engineering challenges. Even if chronic sensing barriers are overcome, achieving regulatory certification and clinical approval for an autonomous physiological feedback controller remains a complex regulatory bottleneck, demanding rigorous validation to meet the critical fail-safe standards of life-support systems. Beyond these engineering and regulatory hurdles, the algorithms driving these controllers must also address profound physiological complexities. Recent models of heart rate and blood pressure variability demonstrate that dynamic structural alterations directly modulate these signals, independent of autonomic tone [47]. Therefore, to ensure clinical safety, future hybrid VAD controllers will require multimodal measurements capable of decoupling mechanical from autonomic contributions, particularly during dynamic challenges such as postural transitions.

## 6. CONCLUSION

The ISA-VAD physiological controller demonstrated consistency and versatility across multiple simulated scenarios, distinguished by its dual-function logic and wide operational range. The in-silico results confirmed that all RQs were successfully addressed.

Regarding RQ1, the ISA-VAD operated under variable-speed conditions ($5.4–9 \cdot 10^3$ r/min), highlighting its adaptive capacity to diverse hemodynamic scenarios, in contrast to fixed-speed controllers. For RQ2, the system dynamically responded to physiological demands by adjusting its rotational speed according to changes in preload, afterload, and HR, thereby ensuring personalized support for the user. Regarding RQ3, the safety module effectively mitigated adverse events by preventing suction and backflow even under high-demand conditions. Finally, with respect to RQ4, measurable clinical improvements were observed, including a 50% reduction in $MVO_2$, an increase in ejection fraction (from 24% to 61%), and an improvement in cardiac output (from 5 $L \cdot min^{-1}$ to 7 $L \cdot min^{-1}$), indicating significant hemodynamic benefits.

The primary RQ was also fulfilled, as the in-silico results demonstrated that the ISA-VAD physiological controller could regulate operational variables while remaining free of suction and backflow. Furthermore, it aligns with state-of-the-art approaches in adaptive control and AI, while maintaining a modular and interpretable hierarchical architecture that favors practical implementation and compatibility with existing commercial controllers. This represents a significant advance in the development of physiological control strategies for VADs considering that the system was evaluated using physiological metrics that are unprecedented in the literature.

These findings reinforce the potential of the ISA-VAD physiological controller to dynamically optimize hemodynamics, reduce the risk of complications, and establish a solid foundation for future in-vitro experiments to validate and expand these results.

## ACKNOWLEDGEMENTS

The authors would like to thank the financial support of the São Paulo State Foundation (FAPESP, Grant 2012/50283-6), FINEP (Grant 01.14.0177.00), and the National Council for Research and Development (CNPq).

## ETHICS STATEMENT

None.